\documentclass[preprint]{elsarticle}

\usepackage{amssymb}
\usepackage{graphicx}
\usepackage{tabularx}
\usepackage{bm}
\usepackage{array}
\usepackage{epsfig}
\usepackage{graphicx}
\usepackage[ansinew]{inputenc}

\journal{Journal of Magnetism and Magnetic Materials}

\begin{document}

\begin{frontmatter}

\title{Ba-substitution induced evolution of structural and magnetic properties of La$_{2-x}$Ba$_x$CoIrO$_6$ double perovskites}

\author[AA]{C. A. S. Vieira}, \author[BB]{B. J. Santos}, \author[BB]{J. G. Duque}, \author[CC]{E. M. Bittar}, \author[AA]{L. Bufai\c{c}al\corref{Bufaical}},\ead{lbufaical@ufg.br}

\address[AA]{Instituto de F\'{i}sica, Universidade Federal de Goi\'{a}s, 74001-970 , Goi\^{a}nia, GO, Brazil}

\address[BB]{Programa de P\'{o}s-Gradua\c{c}\~{a}o em F\'{i}sica, Campus Prof. Jos\'{e} Alu\'{i}sio de Campos, UFS, 49100-000 S\~{a}o Crist\'{o}v\~{a}o, SE, Brazil}

\address[CC]{Centro Brasileiro de Pesquisas F\'{\i}sicas, Rua Dr. Xavier Sigaud 150, 22290-180, Rio de Janeiro, RJ, Brazil}

\cortext[Bufaical]{Corresponding author}

\begin{abstract}

The Iridium-based oxides are the subject of great recent interest due to the non-conventional physics that may emerge from the strong spin-orbit coupling present in 5$d$ ions. Here, we explore the coupling between Ir and Co in the La$_{2-x}$Ba$_x$CoIrO$6$ perovskites ($x$ = 0, 0.5, 0.75 and 1.0), where the structural, electronic, and magnetic properties of the series are investigated by means of x-ray powder diffraction and magnetometry. The system's crystal structure evolves from the monoclinic $P2{1}/n$ to the triclinic $I\overline{1}$ space group as the Ba concentration increases. Measurements of magnetization revealed ferrimagnetic behavior in $x$ = 0, 0.5 and 0.75 compounds, possibly resulting from antiferromagnetic coupling between Co$^{2+/3+}$ and Ir$^{4+}$. In contrast, for $x$ = 1.0 a clear collinear antiferromagnetic character is observed for the Co$^{2+}$ ions, resulting from the quenching of the Ir$^{5+}$ magnetic moment. The evolution of the magnetic properties of the series is discussed in terms of the structural and electronic changes, as well as the spin-orbit coupling in Ir.

\end{abstract}

\begin{keyword}
Iridium; Cobalt; Spin-orbit coupling; Double-perovskite
\end{keyword}

\end{frontmatter}

\section{Introduction}

The strong spin-orbit coupling (SOC) underlying the physics of 5$d$-based oxides can lead to a variety of exotic phenomena. For iridates specifically, the interplay between SOC, electron correlation, and crystal field effects creates a rich landscape for unconventional quantum states, such as topological Mott insulators \cite{Pesin}, Weyl semimetals \cite{Wan}, spin-liquid candidates \cite{Okamoto}, and Kitaev materials \cite{Hickey}. Among these, double perovskites (DP) with the general formula A$_2$BB'O$_6$ provide a feasible framework to modulate magnetic interactions through chemical substitution, due to the structure's flexibility to accommodate various rare-earth and alkaline-earth ions at the A site, and transition-metal (TM) ions at the B/B' sites \cite{Thakur,Sami,Serrate}.

Particularly, 3$d$-5$d$ DPs, where a 3$d$ magnetic ion couples with a 5$d$ ion with strong SOC, have attracted attention due to their interesting magnetic ground states \cite{Alff,Cao}. La$_2$CoIrO$_6$ is a paradigmatic example of a 3$d$-5$d$ DP, on which the antiferromagnetic (AFM) coupling between Co$^{2+}$ (3$d^7$) and Ir$^{4+}$ (5$d^5$) leads to a ferrimagnetic (FIM) and insulating system that undergoes magnetodielectric effect below $T_C$ \cite{Song}. When La$^{3+}$ is partially replaced by Ca$^{2+}$, the changes imposed in the Co/Ir oxidation states leads to competing magnetic interactions that result in spin-glass behavior \cite{PRB2020}. The similar ionic radii of Ca$^{2+}$ and La$^{3+}$ in XII coordination (1.34 and 1.36 \AA, respectively) \cite{Shannon} leads to minimal structural changes with doping in this series, with the monoclinic $P2_{1}/n$ space group being maintained along the whole series \cite{JSSC2015}. Therefore, the magnetic evolution of the La$_{2-x}$Ca$_x$CoIrO$_6$ system may be primarily associated with the electronic changes.

For Sr$^{2+}$ substitution at the La$^{3+}$ site in La$_{2-x}$Sr$_x$CoIrO$_6$, a non-collinear magnetic structure is observed that is dramatically affected by the Sr content \cite{Narayanan,Kolchinskaya}. In this case, albeit the impact of Sr$^{2+}$ in the crystal structure is higher than that of Ca$^{2+}$ due to the former's larger ionic radius (1.44 \AA in XII coordination \cite{Shannon}), the $P2_{1}/n$ symmetry is preserved up to 50\% of substitution. To access more pronounced crystallographic changes, the use of even more discrepant elements at the A-site is necessary.

In this context, Ba-doping in the La-site provides an effective way to induce structural changes due to the large ionic radius of Ba$^{2+}$ (1.61 $\AA$ in XII coordination \cite{Shannon}). Furthermore, Ba-doping changes the valence state balance, potentially altering the oxidation states of Co and Ir ions, thus affecting the magnetic interactions. In this work, we systematically study the structural, electronic, and magnetic evolution of La$_{2-x}$Ba$_x$CoIrO$_6$ ($x$ = 0, 0.5, 0.75, 1.0) perovskites. By employing x-ray powder diffraction (XRD) and magnetization measurements, we explore how Ba-substitution influences the crystal structure, electronic configuration, and magnetic ground state of this series. The observed phenomena are discussed in terms of lattice distortions, valence state modifications, and the pivotal role of SOC in modulating the magnetic properties of these compounds.

\section{Experimental Methods}

The polycrystalline samples of the La$_{2-x}$Ba$_x$CoIrO$_6$ series ($x$ = 0, 0.5, 0.75 and 1.0) here investigated were synthesized by conventional solid-state reaction, in an air atmosphere. Stoichiometric amounts of La$_{2}$O$_{3}$ (purity $\geq$ 99.9\%), BaCO$_3$ (purity $\geq$ 99\%), Co$_{3}$O$_{4}$ (purity $\geq$ 99.5\%)and metallic Ir in powder form (purity $\geq$ 99.9\%), all from Sigma Aldrich, were mixed and heated at $900^{\circ}$C for 24 hours in an air atmosphere. Later, the samples were re-grinded before a second step at $1100^{\circ}$C for 48 hours. Finally, each material was grinded, pressed into pellets, and heated at $1200^{\circ}$C for an additional 24 hours. Attempts to produce the $x$ = 0.25 concentration using the same synthesis route did not succeed, resulting in the formation of impurity phases.

X-ray powder diffraction (XRD) data were collected at room temperature using a Cu $K_{\alpha}$ radiation operating at 40 kV and 40 mA at Centro Brasileiro de Pesquisas F\'{\i}sicas (CBPF), Brazil. The XRD data was carried over the angular range of $10^{\circ}\leq2\theta\leq90^{\circ}$, with a 2${\theta}$ step size of 0.01$^{\circ}$. Rietveld refinement was performed using GSAS software and its graphical interface program \cite{GSAS}. The refinements were conducted assuming a Pseudo-Voigt function for the peak-profiles, and a 8-terms shifted Chebyschev function for the background. The La/Ba isotropic atomic displacement parameters (U$_{iso}$) and atomic positions were constrained to be the same, as well as the U$_{iso}$ of the oxygen ions, in order to avoid divergence during the refinements.

The dc magnetic measurements were carried out using a Quantum Design PPMS magnetometer at CBPF. Magnetization as a function of temperature [$M(T)$] measurements were performed in both zero-field-cooled (ZFC) and field-cooled (FC) modes. Magnetization as a function of applied magnetic field [$M(H)$] curves were carried out after the ZFC protocol.

\section{Results and Discussion}

\begin{figure}
\begin{center}
\includegraphics[width=0.8 \textwidth]{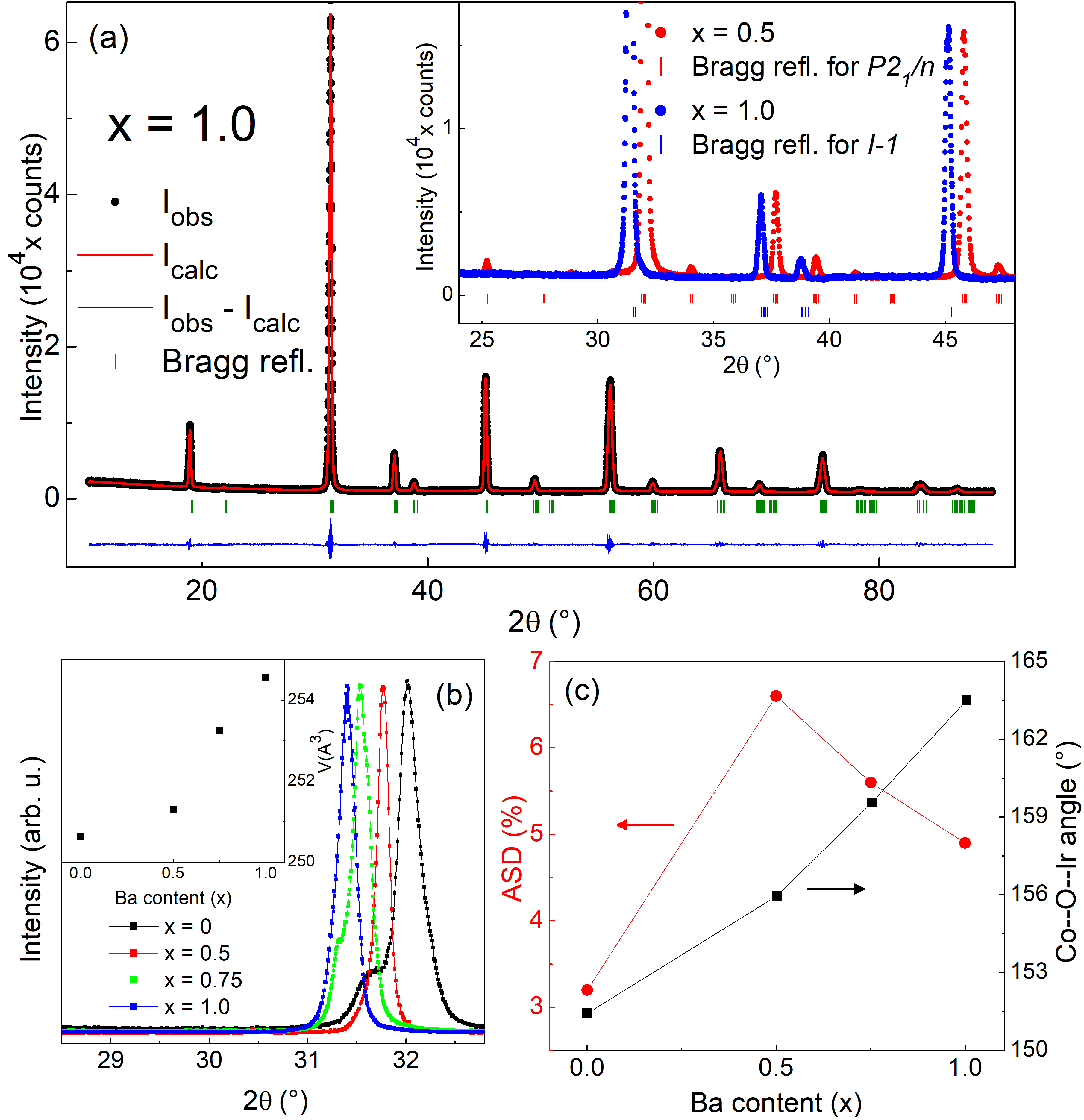}
\end{center}
\caption{(a) Rietveld refinement fitting of the $x$ = 1.0 crystal structure from XRD. The inset compares the XRD patterns of the $x$ = 0.5 and 1.0 samples, where the vertical lines represent the Bragg reflections for the $P2_{1}/n$ and $I\overline{1}$ space groups, highlighting the distinct diffraction peaks. (b) Main diffraction peaks for $x$ = 0, 0.5, 0.75 and 1.0, where the inset shows the evolution of the unit cell volume. (c) Co/Ir ASD and $\langle$Co-O-Ir$\rangle$ bond angle as a function of Ba content.}
\label{Fig_XRD}
\end{figure}

Fig. \ref{Fig_XRD}(a) shows the Rietveld refinement fitting of the XRD pattern of the $x$ = 1.0 sample as a representative of the whole series, for which similar reliability factors were achieved, as depicted in Table \ref{T1}. The Rietveld refinement fittings of the other samples here investigated are depicted in the Supplementary Material, together with the atomic coordinates and other information regarding the crystal structures.

In agreement with previous reports, the crystal structures of both $x$ = 0 and 0.5 could be well refined within the monoclinic $P2_{1}/n$ space group \cite{Song,PRB2020,PRB2023}, signaling that a 25\% Ba substitution at the La site is insufficient to induce a structural transition. For $x$ = 1.0, however, attempts to refine the crystal structure in this same monoclinic symmetry did not succeed. Instead, a good match was achieved with the triclinic $I\overline{1}$ space group. The inset of Fig. \ref{Fig_XRD}(a) compares stretches of the $x$ = 0.5 and 1.0 patterns, endorsing the structural transition. For $x$ = 0.75 the crystal structure could not be well refined when a single phase belonging to $P2_{1}/n$ or to $I\overline{1}$ space group was considered individually. But a good match was achieved when both space groups were considered simultaneously. The refinement indicated about 60\% and 40\% in mass of the $P2_{1}/n$ and $I\overline{1}$ phases, respectively, therefore evidencing that this sample represents an intermediate case between the $x$ = 0.5 and 1.0 compounds.

Fig. \ref{Fig_XRD}(b) compares selected Bragg reflections for the samples of interest, where a systematic shift toward smaller 2${\theta}$ angles with increasing Ba content is evidenced. As depicted in the inset, this reflects the increase in the unit cell volume ($V$), related to the fact that the Ba$^{2+}$ ionic radius is significantly larger than that of La$^{3+}$ (respectively 1.61 and 1.36 \AA, in XII coordination) \cite{Shannon}. It should be noted that the $V$ expansion is remarkably larger from $x$ = 0.5 to 1.0 than from $x$ = 0 to 0.5. This is most likely related to the prominent decrease of the average B-site radius from $x$ = 0 to 0.5 since, as will be discussed in the context of the magnetic properties, the 25\% Ba$^{2+}$ for La$^{3+}$ substitution primarily increases the Co formal valence from Co$^{2+}$ to Co$^{2.5+}$ (the Co$^{2+}$ and Co$^{3+}$ radii are respectively 0.745 and 0.61 \AA, in high spin state) \cite{Shannon}. On the other hand, for further Ba-doping there is a tendency to return to the Co$^{2+}$ configuration, while Ir valence tends to increase (Ir$^{4+}$ and Ir$^{5+}$ radii are 0.625 and 0.57 \AA, respectively) \cite{Shannon}. The $V$ expansion with increasing Ba content is also manifested in the increase of the average Co-O-Ir bond angle, Fig. \ref{Fig_XRD}(c). This was expected, since the B-O-B' angles usually increase with $V$ in DPs \cite{Sami,Serrate}. 

\begin{table}
\caption{Main lattice parameters and reliability factors obtained from the Rietveld refinements. The numbers in parentheses represents the standard uncertainties in the last digits.}
\label{T1}
\begin{tabular}{c|c|c|cc|c}
\hline \hline
Sample ($x$) & 0 & 0.5 & 0.75 & 0.75 & 1.0 \\

\hline

SG & $P2_{1}/n$ & $P2_{1}/n$ & $P2_{1}/n$ (60\%) & $I\overline{1}$ (40\%) & $I\overline{1}$ \\

$a$ (\AA) & 5.57789(16) & 5.63357(14) & 5.61075(9) & 5.63202(7) & 5.65156(8)\\

$b$ (\AA) & 5.67845(15) & 5.61535(14) & 5.66650(10) & 5.63058(9) & 5.63271(10)\\

$c$ (\AA) & 7.91281(19) & 7.94361(21) & 7.96309(8) & 7.99013(13) & 7.99710(10) \\

$V$ (\AA$^{3}$) & 250.629(14) & 251.292(5) & 253.172(9) & 253.360(5) & 254.568(8) \\

$\alpha$ ($^{\circ}$) & 90 & 90 & 90 & 89.708(1) & 89.631(2) \\

$\beta$ ($^{\circ}$) & 89.956(5) & 90.036(4) & 89.852(1) & 90.508(1) & 90.280(2)\\

$\gamma$ ($^{\circ}$) & 90 & 90 & 90 & 90.390(1) & 90.061(2) \\

ASD (\%) & 4.4 & 7.5 & 4.9 & 6.6 & 4.9 \\

\hline

$\langle$Co-O$\rangle$ (\AA) & 2.0088(12) &	 2.0090(11) & 2.0152(6) & 2.0049(7) & 1.9931(13) \\

$\langle$Ir-O$\rangle$ (\AA) & 2.0959(12) &	2.0555(11) & 2.0388(6) & 2.0498(7) & 2.0449(13) \\

$\langle$Co-O-Ir$\rangle$ ($^{\circ}$) & 151.456(17) & 155.946(16) & 159.069(12) & 160.318(13) & 163.503(16) \\

\hline

$R_{wp}$ (\%) & 6.3 & 4.9 & 4.2 & 4.2 & 4.7 \\

$R_{p}$ (\%) & 4.8 & 3.5 & 3.2 & 3.2 & 3.7 \\

\hline \hline

\end{tabular}
\end{table}

Importantly, the anti-site disorder (ASD) at the Co and Ir sites, extracted from the Rietveld refinements, does not evolve monotonically. As Fig. \ref{Fig_XRD}(c) shows, it initially increases from $x$ = 0 to 0.5, followed by a systematic decrease for larger Ba-doping. The ASD is known to be ruled by the differences between the B/B' charges and ionic radii, as well as by the unit cell volume \cite{Alam}. Therefore, the changes observed for the ASD are, again, directly related to the evolution of the crystal structure and the Co/Ir electronic configurations. For $x$ = 0, one has mainly Co$^{2+}$ and Ir$^{4+}$, whose charges and radii differ significantly \cite{PRB2020}. Thus, a relatively small ASD is noticed. For $x$ = 0.5, as will be discussed in the context of magnetic properties, the introduction of Ba$^{2+}$ at the La$^{3+}$ site most likely acts to increase the Co valence state, while Ir remains nearly tetravalent. The presence of 50\% of Co$^{3+}$ makes the charges and radii of the transition metals (TM) more alike. Furthermore, the $V$ expansion facilitates the permutation between the TMs, resulting in a pronounced increase in the ASD. On the other hand, for Ba-doping larger than 25\% there is a tendency to return to Co$^{2+}$ while the Ir valence increases, resulting in larger differences between the Co/Ir charges and radii that lead to the decrement the ASD. However, $V$ systematically enhances with Ba-content. From these joint effects it results that, although the ASD decreases from $x$ = 0.5 to 1.0, it does not achieve the same small value as that found for $x$ = 0. As we shall see next, the structural changes observed along the series directly impact the system's magnetic properties.

\begin{figure}
\begin{center}
\includegraphics[width=0.8 \textwidth]{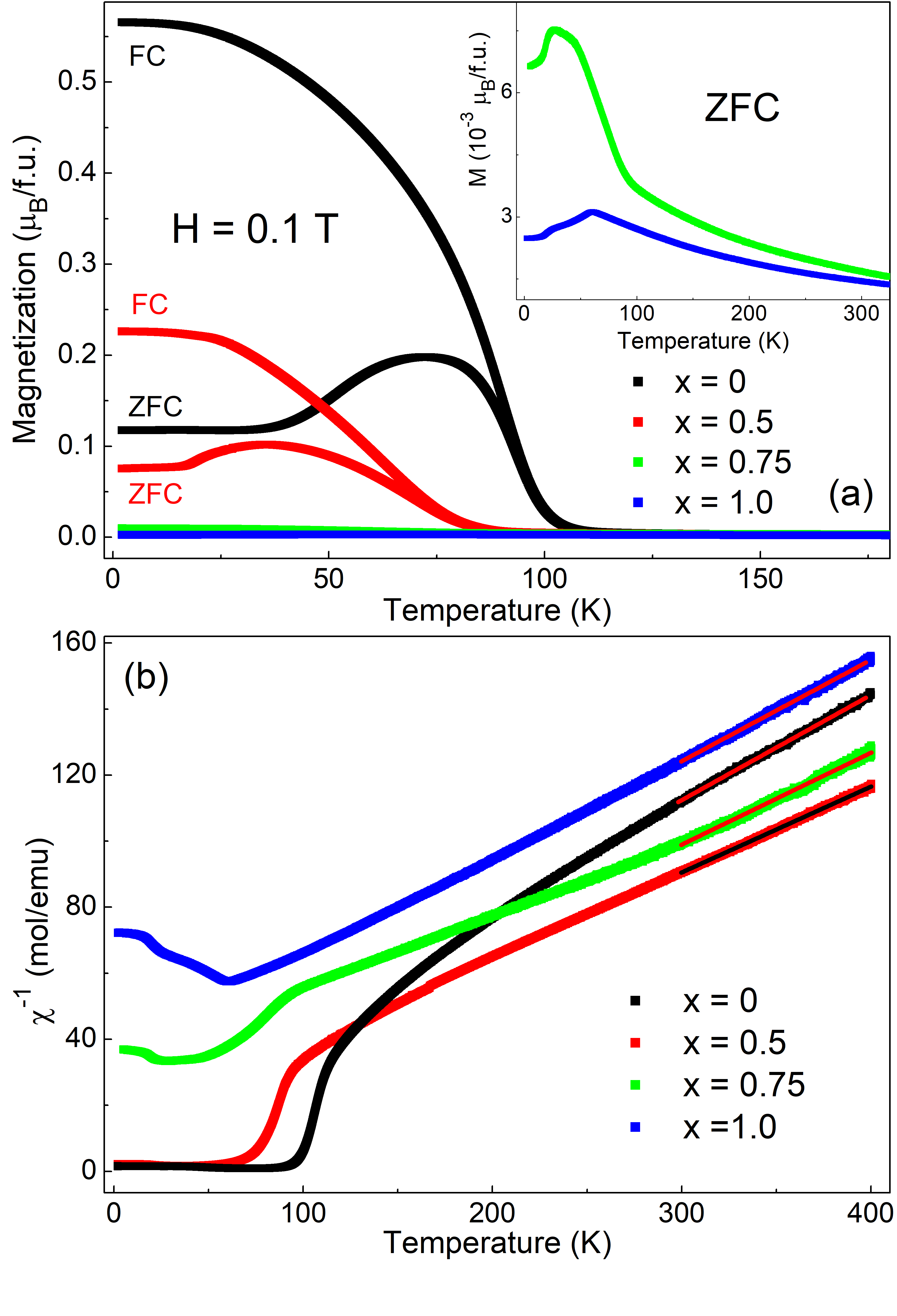}
\end{center}
\caption{(a) ZFC-FC $M(T)$ curves measured with $H$ = 0.1 T. The inset shows a magnified view of the low temperature region of $x$ = 1.0, highlighting its $T_N$. (b) $\chi^{-1}$ curves, where the straight lines represent the best fits with the CW law.}
\label{Fig_MxT}
\end{figure}

The ZFC-FC $M(T)$ measurements, performed with a magnetic field $H$ = 0.1 T, are depicted in Fig. \ref{Fig_MxT}. The FC curves of $x$ = 0 and 0.5 are typical of ferromagnetic (FM) or FIM compounds. The magnitudes of the magnetization at low temperatures, however, are much smaller than expected for a FM, indicating a FIM character that results from antiferromagnetic (AFM) coupling between Co and Ir. The magnetic ordering temperature of $x$ = 0, $T_C$ $\simeq$ 94 K, agrees with previous reports, being associated with the onset of AFM coupling between Co$^{2+}$ and Ir$^{4+}$ \cite{Song,JSSC2015}. For $x$ = 0.5, $T_C$ decreases to approximately 70 K. It must be noticed that the magnitude of the magnetization is also reduced when compared with the pristine compound. Similar behavior was observed in the La$_{2-x}$Ca$_x$CoIrO$6$ and La$_{2-x}$Sr$_x$CoIrO$_6$ series, for which the magnetization progressively decreases with increasing the content of alkaline-earth ion \cite{PRB2020,Narayanan}. Here, such evolution must be related to changes in the magnetic structure and/or to the increased ASD at $x$ = 0.5 that hampers the percolation of  exchange interactions due to the increased presence of competing magnetic phases and frustration.

For $x$ = 0.75 and 1.0, however, the dramatic reduction of the magnetization cannot be accounted for by the ASD. The inset of Fig. \ref{Fig_MxT}(b) shows magnified views of the $M(T)$ curves for these compounds, where a cusp typical of AFM systems is noticed at $T_N$ $\simeq$ 61 K for $x$ = 1.0, followed by a broader peak at $\sim$25 K. The low temperature anomaly is possibly related to the onset of glassy magnetic behavior, or to the presence of short-range Co-Co/Ir-Ir couplings brought by the ASD. In the latter case, the short-range correlations may be also present in the $x$ = 0 and 0.5 compounds, but they would be masked by the magnetization from the major FIM coupling. The use of additional techniques, such as AC magnetic susceptibility, is necessary to unravel this issue. Interestingly, the ZFC curve for $x$ = 0.75 seems to present both characteristics of FIM and AFM couplings. It shows a magnetic transition similar to that of FM/FIM compounds at $\sim$66 K followed by a cusp similar to that of AFM systems at $\sim$26 K. Details about the magnetic properties of this compound will be addressed below.

The fits of the reciprocal magnetic susceptibility ($\chi^{-1}$) curves with the Curie-Weiss (CW) law are displayed in Fig. \ref{Fig_MxT}(c). For $x$ = 0, this yields a CW temperature $\theta_{CW}$ = -53 K and an effective magnetic moment $\mu_{eff}$ = 5.0 $\mu_B$/f.u.. The negative $\theta_{CW}$ confirms the predominance of AFM coupling, corroborating the FIM scenario. The experimental effective moment can be compared with the expected value calculated with the equation usually adopted for systems presenting two or more magnetic ions
\begin{equation}
\mu_{calc} = \sqrt{{\mu_1}^2 + {\mu_2}^2 + {\mu_3}^2...}. \label{EqMU}
\end{equation}
Assuming the standard individual moments of Co$^{2+}$ ($\mu_{Co^{2+}}$ = 4.8 $\mu_B$) and Ir$^{4+}$ ($\mu_{Ir^{4+}}$ = 1.7 $\mu_B$) \cite{Ashcroft}, one obtains $\mu_{calc}$ = 5.1 $\mu_B$/f.u., fairly close to the experiment.

\begin{table}
\caption{Main results obtained from the $M(T)$ and $M(H)$ curves.}
\label{T2}
\begin{tabular}{c|cccc}
\hline \hline
Sample ($x$) & 0 &  0.5 &  0.75  &  1.0 \\

\hline

$T_C$/$T_N$ (K) & 94 & 70 & 66 & 61 \\

$\theta_{CW}$ (K) & -53 & -48 & -56 & -104 \\

$\mu_{eff}$ ($\mu_B$/f.u.) & 5.0 & 5.5 & 5.3 & 5.1 \\

$\mu_{calc}$ ($\mu_B$/f.u.) & 5.1 & 5.4 & 5.3 & 4.8 \\

$H_C$ (T) & 1.8 & 1.9 & 0.15 & - \\

$M_R$ ($\mu_B$/f.u.) & 0.669 & 0.253 & 0.007 & - \\

\hline \hline

\end{tabular}
\end{table}

For $x$ = 0.5, the CW fit yields $\theta_{CW}$ = -48 K and $\mu_{eff}$ = 5.5 $\mu_B$/f.u.. The enhancement of $\mu_{eff}$ with respect to $x$ = 0 gives further evidence that the the 25\% of Ba$^{2+}$ to La$^{3+}$ substitution leads to the increase of the Co oxidation state to Co$^{2.5+}$. If, instead, Ir valence would been enhanced to +4.5 one should expect the decrease of $\mu_{eff}$, since it is known that the large SOC further lifts the degeneracy of the $t_{2g}$ orbitals in this 5$d$ ion, in a way that Ir$^{5+}$ is expected to have a zero net magnetic moment \cite{PRB2020,Kim,Manna,Ray,Pramanik}. Here we remind that this scenario is endorsed by the increased ASD, caused by the greater similarity between the Co/Ir valences and radii. Adopting the standard moments for Co$^{2+}$, Ir$^{4+}$ and Co$^{3+}$ ($\mu_{Co^{3+}}$ = 5.4 $\mu_B$) \cite{Ashcroft} in Eq. \ref{EqMU}, we get $\mu_{calc}$ = 5.4 $\mu_B$/f.u., also close to the experimental value.

For $x$ = 0.75, the CW fit yields $\theta_{CW}$ = -56 K, while $\mu_{eff}$ decreases to 5.3 $\mu_B$/f.u.. Such a decrease of $\mu_{eff}$ with respect to the value found for $x$ = 0.5 cannot be explained by a further increase in the Co valence, otherwise the effective moment would increase. Instead, the increase of the Ir oxidation state is most likely, also agreeing with the decreased ASD. Assuming in Eq. \ref{EqMU} that the further Ba-doping leads to Ir$^{4.25+}$ configuration, while the Co$^{2.5+}$ formal valence remains unchanged, we have $\mu_{calc}$ = 5.3 $\mu_B$/f.u., matching with the experimentally observed decrease of $\mu_{eff}$.

For $x$ = 1.0, the effective moment further reduces to 5.1 $\mu_B$/f.u.. Two possible scenarios can be invoked to account for this: (i) the Ir oxidation states increases to +4.5 while the Co$^{2.5+}$ mixed valence is maintained; (ii) Ir is fully oxidized to pentavalent state while Co is reduced to the bivalent state. Eq. \ref{EqMU} yields respectively $\mu_{calc}$ = 5.2 and 4.8 $\mu_B$/f.u. for (i) and (ii), matching with the decrease of $\mu_{eff}$ experimentally observed. Although the paramagnetic moment calculated within picture (i) is closer to the experiment, we believe that the later scenario is most likely, since the stabilization of Ir$^{5+}$ have been confirmed for several LaABIrO$_6$ (A = Ca, Sr, Ba; B = 3$d$ TM ions) perovskites \cite{PRB2020,Ray,Jansen,JMMM2023,Jansen2}. Furthermore, the line-shape of the $M(T)$ curves observed for this sample are characteristic of a pure AFM system, \textit{i.e.} typical of a single magnetic ion, whereas a FM- or FIM-like curve should be expected for a system presenting two or more magnetic ions. Here, the difference between the calculated and experimental moments could be due to a significant orbital contribution to the Co moment resulting from a weak crystal field, as suggested by the large unit cell volume \cite{Song,PRB2020,Raveau}. In any case, the possible presence of small amounts of Co$^{3+}$/Ir$^{4+}$ cannot be ruled out.

Regarding the overall results of the CW fits, the $\theta_{CW}$ values obtained differ from the $T_C$/$T_N$ observed. Despite the linear character of the curves in the PM region of the $\chi^{-1}$ plots and the correspondingly good fits achieved, the discrepancies could be related to deviations from Curie-Weiss (CW) behavior, as previously reported for similar Ir-based DPs \cite{Cao2,Jansen,JMMM2023}. Nevertheless, the negative $\theta_{CW}$ values obtained for all compounds endorse the scenario of AFM coupling between Co and Ir in $x$ = 0, 0.5 and 0.75, and between the Co moments in $x$ = 1.0.

\begin{figure}
\begin{center}
\includegraphics[width=0.7 \textwidth]{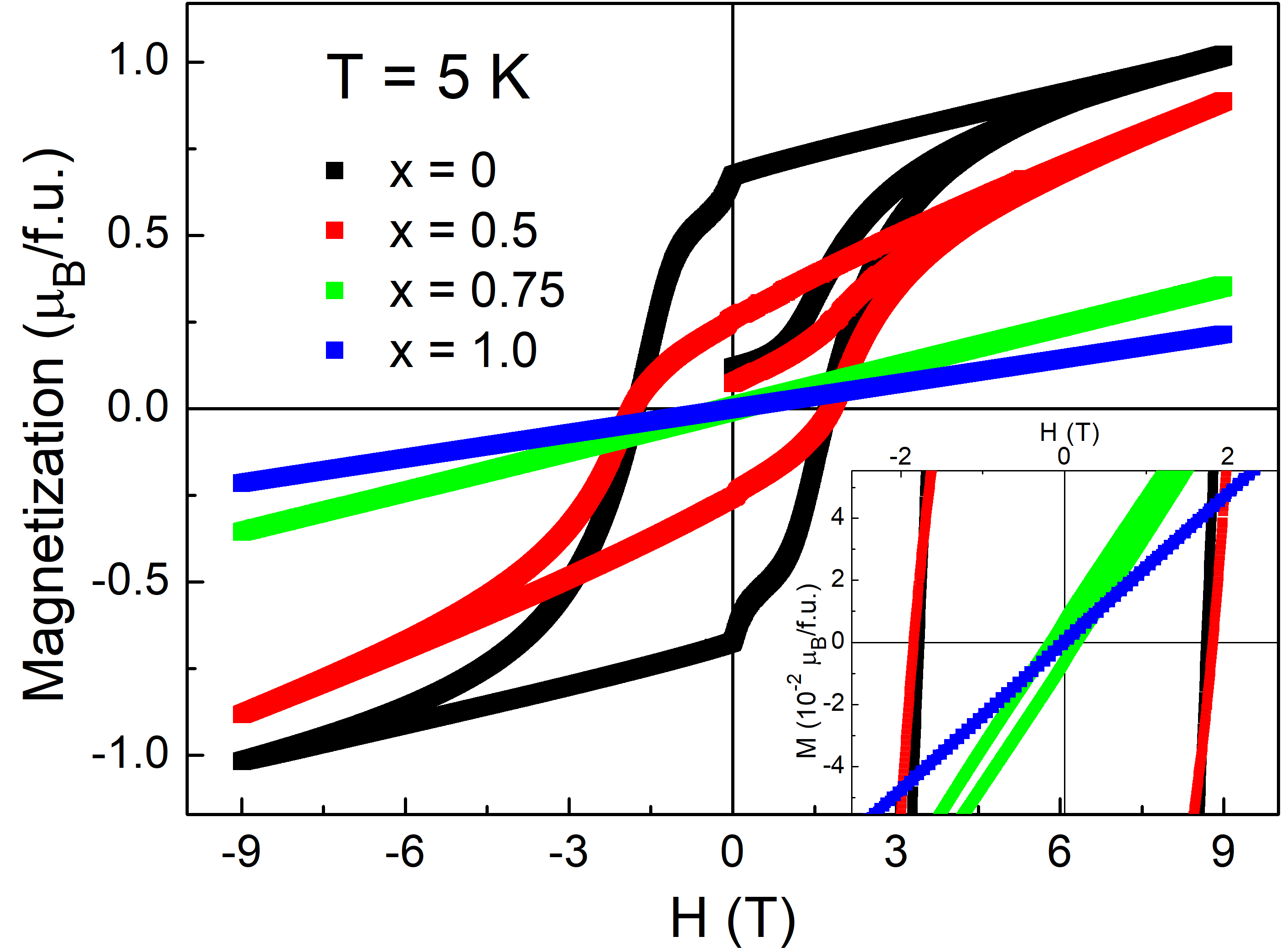}
\end{center}
\caption{$M(H)$ curves measured at 5 K. The inset shows a magnified view of the low field region.}
\label{Fig_MxH}
\end{figure}

The $M(H)$ curves are depicted in Fig. \ref{Fig_MxH}. The loops of $x$ = 0 and 0.5 are typical of FM or FIM polycrystals, with an S-shaped form, large remanent magnetizations ($M_R$) and coercive fields ($H_C$) of the order of 2 T. But again, the magnetization values are much smaller than expected for FM coupling between Co and Ir. Furthermore, the magnetization does not saturate even at $H$ = 9 T, continuing to increase linearly at high fields. These features further suggest a FIM character resulting from AFM coupling between Co and Ir. The decreased magnetization of $x$ = 0.5 with respect to $x$ = 0 may be related to the increased ASD in the former compound. Indeed, the larger slope of the $x$ = 0.5 curve at high fields suggests an augmented contribution of short-range AFM couplings and/or paramagnetic (PM) contributions brought by the ASD. Alternatively, the magnetic evolution observed from $x$ = 0 to 0.5 could be related to changes in presumably non-collinear spin superstructures of Co and Ir sites, as suggested for La$_{2-x}$Sr$_x$CoIrO$_6$ \cite{Narayanan}.

The $M(H)$ curve of $x$ = 0.75 lacks the S-shaped form, and the magnetization is significantly reduced with respect to that of $x$ = 0 and 0.5. But non-negligible $H_C$ and $M_R$ are observed, suggesting the presence of FIM coupling. For $x$ = 1.0, however, the $M(H)$ curve is typical of AFM systems, exhibiting negligible $M_R$ and $H_C$. This endorses the scenario of a purely AFM ordering on the Co superstructure for this compound, due to the zero net magnetic moment of Ir$^{5+}$ caused by its strong SOC that lifts the $t_{2g}$ degeneracy of this 5$d^{4}$ ion into a fully occupied $j_{3/2}$ quadruplet and an empty $j_{1/2}$ doublet. Interestingly, both the $M(T)$ and $M(H)$ curves of LaBaCoIrO$_6$ are distinct from the FM-like shapes observed in LaSrCoIrO$_6$ and LaCaCoIrO$_6$ \cite{Narayanan,JMMM2023}. Albeit the Co$^{2+}$ and non-magnetic Ir$^{5+}$ configurations also dominate the latter compounds, the more distorted crystal structures lead to spin canting on the Co moments. In contrast, for the more symmetric LaBaCoIrO$_6$ system, a collinear AFM superstructure is most likely.

\section{Conclusions}

We have investigated the structural, electronic, and magnetic properties of La$_{2-x}$Ba$_x$CoIrO$_6$ double perovskites ($x$ = 0, 0.5, 0.75, 1.0). Our results reveal a progressive structural evolution from monoclinic to triclinic symmetry upon Ba-doping, accompanied by a suppression of ferrimagnetic order. The anti-site disorder at the Co/Ir sites increases from $x$ = 0 to $x$ = 0.5, this behavior being consistent with the initial increase of the Co oxidation state up to 25\% Ba$^{2+}$ substitution at the La$^{3+}$ site. In contrast, further Ba-doping seems to increase the Ir valence, with this ion reaching the pentavalent state while Co returns to a bivalent configuration for $x$ = 1.0. The structural and electronic evolution explain the changes in the magnetic properties of the series and further support the scenario of large SOC on Ir, where the $x$ = 0, 0.5 and 0.75 samples exhibit FIM behavior resulting from AFM coupling between Co$^{2+/3+}$ and Ir$^{4+}$, while for $x$ = 1.0 the presence of non-magnetic Ir$^{5+}$ leads to a typically collinear AFM character at the Co$^{2+}$ superstructure.

\section*{Acknowledgements}

This work was supported by the Brazilian funding agencies: Funda\c{c}\~{a}o de Amparo \`{a}  Pesquisa do Estado de Goi\'{a}s (FAPEG) [No. AUX2024301000048], Funda\c{c}\~{a}o Carlos Chagas Filho de Amparo \`{a} Pesquisa do Estado do Rio de Janeiro (FAPERJ) [Nos. E-26/204.206/2024 and E-26/211.291/2021], and Conselho Nacional de Desenvlovimento Cient\'{\i}fico e Tecnol\'{o}gico (CNPq) [No. 305394/2023-1].

\end{document}